\documentclass[12pt,a4paper]{iopart}

\expandafter\let\csname equation*\endcsname\relax
\expandafter\let\csname endequation*\endcsname\relax 
\usepackage{amsmath}
\usepackage{iopams}
\usepackage{graphics}
\usepackage{graphicx}
\usepackage{psfrag}
\usepackage{color}
\usepackage{colordvi}

\begin{document}

\title{Some ground-state expectation values for the free parafermion $Z(N)$ spin chain}
\author{Zi-Zhong Liu$^1$, Robert A. Henry$^2$, Murray T. Batchelor$^{2,3,1,4}$ and Huan-Qiang Zhou$^1$}
%\affiliation{
\address{
$^1$Centre for Modern Physics, Chongqing University, Chongqing 400044, China\\
$^2$Department of Theoretical Physics, Research School of Physics and Engineering, 
Australian National University, Canberra ACT 2601, Australia\\
$^3$Mathematical Sciences Institute,
Australian National University, Canberra ACT 2601, Australia\\
$^4$ batchelor@cqu.edu.cn
}

\begin{abstract}
We consider the calculation of ground-state expectation values for the non-Hermitian $Z(N)$ spin chain described by free parafermions. 
For $N=2$ the model reduces to the quantum Ising chain in a transverse field with open boundary conditions. 
Use is made of the Hellmann-Feynman theorem to obtain exact results for particular single site and nearest-neighbour 
ground-state expectation values for general $N$ which are valid for sites deep inside the chain. 
These results are tested numerically for $N=3$, along with how they change as a function of distance from the boundary. 
\end{abstract}

%\begin{center}
\footnotetext{
{\em Dedicated to the memory of Vladimir Rittenberg}}
%\end{center}

\maketitle

\section{Introduction}

The Ising model is one of the most widely studied models in physics.
The understanding obtained from the exact solution of the two-dimensional Ising model and its one-dimensional quantum counterpart played a pivotal role in the 
formulation of the theory of phase transitions and critical phenomena \cite{Baxter2007,McCoy2010,M2010,S2011}. 
Although describing interacting classical and quantum spin systems, a feature of the Ising model is that the exact solution is 
described in terms of free fermions.
It is this property of free fermions which allows the relatively straightforward calculation of physical properties, given the 
powerful framework of statistical mechanics. 

In terms of the Pauli matrices $\sigma^x_j$ and $\sigma^z_j$ acting at site $j$ on a chain of length $L$ the hamiltonian of the one-dimensional quantum Ising model in 
a transverse field can be defined by 
\begin{equation}
H_{\mathrm{f}} = -  \sum_{j=1}^{L-1} \sigma^z_j \sigma^z_{j+1} -  \lambda  \sum_{j=1}^L  \sigma^x_j .
\label{ham}
\end{equation}
The field variable $\lambda$ plays a temperature-like role, with a quantum phase transition at $\lambda=\lambda_c=1$ 
between ferromagnetic ($\lambda<1$) and paramagnetic ($\lambda>1$) phases \cite{S2011}.
Here we have imposed open boundary conditions, for reasons which will become apparent.

The quantum Ising chain belongs to several families of exactly solved $N$-state quantum models. 
Such a more general model of interest here is the $Z(N)$ model defined by the hamiltonian 
\begin{equation}
H_{\mathrm{pf}} = -  \sum_{j=1}^{L-1} \sigma_j \sigma_{j+1}^{\dagger} -  \lambda  \sum_{j=1}^L  \tau_j , 
\label{pf}
\end{equation}
where $\sigma_j$ and $\tau_j$ are $Z(N)$ operators defined in matrix form by 
\begin{eqnarray}
\sigma_j &=& I \otimes I \otimes  \cdots \otimes I \otimes \sigma \otimes I \otimes \cdots \otimes I,\\
\tau_j &=& I \otimes I \otimes  \cdots \otimes I \otimes \tau \otimes I \otimes \cdots \otimes I.
\end{eqnarray}
Here the identity $I$, $\sigma$ and $\tau$ are $N \times N$ matrices, with $\sigma$ and $\tau$ in position $j$. 
The clock and shift matrices $\sigma$ and $\tau$ have components 
\begin{equation}
\sigma_{m,n} =  \omega^{m-1} \delta_{m,n}, \quad \tau_{m,n} = \delta_{m,n+1}, 
\end{equation}
with $\omega = {\mathrm e}^{2\pi {\mathrm i}/N}$ and $\tau_{m,N} = \delta_{m,1}$. 
For $N=2$ these are the Pauli matrices $\sigma^z$ and $\sigma^x$, with hamiltonian (\ref{pf}) reducing to hamiltonian (\ref{ham}). 
For general $N$ the clock and shift matrices satisfy the relations 
\begin{equation}
\sigma \tau = \omega \tau \sigma,
\quad \tau^\dagger = \tau^{N-1}, \quad \sigma^\dagger = \sigma^{N-1},
\end{equation}
with $\sigma^N = \tau^N = I$.

The $Z(N)$ model defined by hamiltonian (\ref{pf}) was introduced by Baxter \cite{Baxter1989a,Baxter1989b,Baxter2004}. 
It follows from a version of the $\tau_2$ model, also known as the Bazhanov-Stroganov model~\cite{BS}, 
a two-dimensional classical chiral spin model connecting 
the six-vertex model and the chiral Potts model in such a way that the chiral Potts model 
can be viewed as a descendent of the six-vertex model~\cite{BS,Perk2016}.

It is important to note that the $Z(N)$ hamiltonian (\ref{pf}) is non-Hermitian for $N \ge 3$, with a complex eigenspectrum. 
With the addition of the Hermitian conjugate term this model has been studied recently in the context of parafermionic edge modes \cite{hcon1,hcon2}, mostly for $N=3$. 
The non-Hermitian $Z(N)$ hamiltonian (\ref{pf}) has been solved exactly only for open boundary conditions. 
The energy eigenspectrum displays the remarkable property of free parafermions \cite{Fendley2014}.
This free parafermionic structure has also been confirmed in the related
$\tau_2$ model with open boundaries~\cite{Baxter2014,YP2014,YP2016}.
The $Z(N)$ hamiltonian (\ref{pf}) and the related $\tau_2$ model are the 
only known models with an entire spectrum described precisely by free parafermions.

Relatively little is known so far about the physical properties of free parafermions, 
compared to free fermions, for which a wealth of information is known.
The free parafermion $Z(N)$ model (\ref{pf}) has been shown \cite{ABL2017,AB2018} to share some critical exponents with the %superintegrable 
other parafermionic $Z(N)$ models. 
For example, the specific heat and anisotropic correlation length exponents, $\alpha = 1 - 2/N$, $\nu_\parallel = 1$ and $\nu_\perp = 2/N$, are the same as for the 
superintegrable $N$-state chiral Potts model \cite{Baxter1989b,AMPT,Perk2016}. 
This specific heat exponent is also shared by the $Z(N)$ Fateev-Zamolodchikov model \cite{FZ}. 
We note also that the free parafermion $Z(N)$ model provides what appears to be the first example of a one-dimensional hamiltonian exhibiting boundary dependent bulk behavior. 
For example, the bulk ground-state energy per site and various critical exponents differ 
between open and periodic boundary conditions \cite{AB2018}.

Our specific interest here is in the nature of spin correlations.  
Spin correlations and related order parameters for the Ising model, in both the classical and quantum formulations, are well understood (see, e.g.,~\cite{McCoy2010}).
Although much work remains to be done for the generalization of the Ising results to the $N$-state chiral Potts model,  
there is a remarkable result for the order parameters $M_k$, $k=1, 2, \ldots, N-1$.
The general result, 
\begin{equation}
M_k = (1-\lambda^2)^{k(N-k)/(2N^2)}, 
\label{spon}
\end{equation}
reduces to the spontaneous magnetization of the Ising model for $N = 2$ and $k = 1$. 
The formula (\ref{spon}) was first conjectured by Albertini, McCoy, Perk and Tang in 1989 \cite{AMPT}, then finally proven by Baxter in 2005 \cite{Baxter2005}.
This result is the culmination of a very fruitful conjectural path beginning with   
Howes, Kadanoff and den Nijs \cite{HKN1983} for $N=3$ and the  
discovery of the $N$-state superintegrable chiral Potts chain by von Gehlen and Rittenberg \cite{vGR1985}.

Here we investigate some ground-state expectation values for the free parafermion $Z(N)$ model. 
We focus on open boundary conditions for which the $Z(N)$ free parafermion model is 
exactly solved and for which exact information is readily obtained for the Ising case.
As a special case, the $N=2$ results are thus of relevance to the study of the more general free parafermion model.
In Section \ref{solution} we summarise the known solution for the eigenvalues and eigenvectors 
of the Ising hamiltonian (\ref{ham}) with open boundary conditions. 
These results are used in Section \ref{results2} to consider various exact spin correlations for the open quantum Ising chain.
Exact results are derived for some ground-state expectation values for the open $Z(N)$ chain in Section \ref{exactN}.
These results are tested numerically in Section \ref{resultsN} for $N=3$. 
A concluding discussion of the results is given in Section \ref{conclude}.

\section{Eigenvectors and spin correlations for the open Ising chain}
\label{solution}

The hamiltonian (\ref{ham}) was solved long ago by Pfeuty \cite{P1970}, closely following the similar solution of the XY model with open 
boundary conditions  \cite{LSM1961}. 
The open quantum Ising chain has been discussed by a number of authors in different contexts, 
including the dynamics (see, e.g., Refs \cite{K2004,CPV2015,F2016} and references therein).
Early work was also done on this model in relation to the connection between finite-size corrections and conformal field theory \cite{CJ1987}.
A pedagogic description of the solution of both models in terms of free fermions has been given by Karevski \cite{K2004}. 
The key ingredient is the Jordan-Wigner transformation, as applied  
in the seminal paper by Lieb, Schultz and Mattis \cite{LSM1961} outlining the solution of the XY model in terms of free fermions.
Indeed the XY and quantum Ising models are closely related and can be treated under the same banner 
in the form of the XYh model (see, e.g., Refs \cite{K2004,F2016} for such a treatment for open boundary conditions).
In this Section we summarise some results of relevance to the calculation of the spin correlations for the open quantum Ising chain defined by hamiltonian (\ref{ham}).

The energy eigenspectrum of hamiltonian (\ref{ham}) has the simple free fermion form 
\begin{equation}
E= \pm \epsilon_1 \pm  \epsilon_2 \pm \cdots \pm  \epsilon_L .
\label{spec}
\end{equation}
This covers all $2^L$ eigenvalues in the energy spectrum for a chain of length $L$.
The quasi energy levels $\epsilon_j$ appearing in (\ref{spec}) are functions of $\lambda$.
They are determined by the eigenvalues of the $L \times L$ matrices $C^\dagger C$ or $C C^\dagger$, where 
\begin{equation}
C=\begin{bmatrix} \lambda &  &  & &\\ 1 &\lambda&&&\\  & 1 & \lambda & &\\  &  & \ddots & \ddots & \\& &  & 1 & \lambda 
\end{bmatrix} , \quad
C^\dagger=\begin{bmatrix} \lambda & 1 &  & &\\  &\lambda&1 &&\\  &  & \ddots & \ddots&\\  &  &  & \lambda & 1 \\& &  &  & \lambda 
\end{bmatrix} . 
\end{equation}
These eigenvalues are given by 
\begin{equation}
\epsilon_{j} = \left( 1 + \lambda^2 + 2 \lambda \cos {k_j} \right)^{1/2}.
\end{equation}
The roots $k_j$, $j=1,\ldots,L$, satisfy the equation 
\begin{equation}
{\sin(L+1) k = -{\lambda}^{-1} \sin Lk}.
\label{eqn}
\end{equation}

The relevant eigenvectors are given in terms of the matrices $C$ and $C^\dagger$ by $C \psi_k = \epsilon_k \phi_k$ and $C^\dagger \phi_k = \epsilon_k \psi_k$.  
Thus
\begin{align}
C C^\dagger \phi_k = \epsilon^2_k \phi_k, \\
C^\dagger  \! C \psi_k = \epsilon^2_k \psi_k .
\end{align}
The eigenvector components are given explicitly by
\begin{align}
&\phi_{k_j}(n) = A_{k_j} \sin(L+1-n) k_j , \\ 
&\psi_{k_j}(n) = -(-1)^j  A_{k_j} \sin n k_j ,
\end{align}
for $n=1,\ldots,L$. The normalization factor $A_{k_j}$ is 
\begin{equation*}
A_{k_j}^2 = \frac{4}{2L+1 - \csc k_j \sin(2L+1) k_j} .
\end{equation*}

Given the eigenvector components, a key ingredient in the calculation of spin correlations is the quantity 
\begin{equation}
G_{mn} = - \sum_k \phi_k(m) \psi_k(n) ,
\end{equation}
in terms of which, for example,  the end-to-end spin correlations are determined by  \cite{LSM1961}
\begin{eqnarray}
\langle \Psi_0 | \sigma^x_1 \sigma^x_L  | \Psi_0 \rangle &=& G_{11} G_{LL} - G_{1L} G_{L1}, \label{XX} \\
\langle \Psi_0 | \sigma^y_1 \sigma^y_L  | \Psi_0 \rangle &=& G_{L1}, \label{YY} \\
\langle \Psi_0 | \sigma^z_1 \sigma^z_L  | \Psi_0 \rangle &=& G_{1L}, \label{ZZ}
\end{eqnarray}
Here $\Psi_0$ is the ground-state eigenvector of hamiltonian ({\ref{ham}). 
It is convenient to define the limits
\begin{equation}
\rho_a = \lim_{L\to\infty} \langle \Psi_0 | \sigma^a_1 \sigma^a_L  | \Psi_0 \rangle,
\end{equation}
with $a=x,y,z$. 

In this way, Pfeuty \cite{P1970} obtained the result 
\begin{equation}
\rho_z   = \begin{cases}
1 - \lambda^2, \quad \lambda < 1\\
0, \qquad \quad \, \lambda \ge 1 \, .
\end{cases} 
\label{ZZinf}
\end{equation}
As noted at the time, this differs from  the result 
\begin{equation}
\rho_z   =  \begin{cases}
(1 - \lambda^2)^{1/4}, \quad \lambda < 1\\
0, \qquad \qquad \quad \,\, \lambda \ge 1 ,
\end{cases}
\label{rhoz}
\end{equation}
obtained for periodic boundary conditions \cite{W1966,M1968,P1970}. %, for which the sites at either end of the chain are nearest neighbors.
To obtain this result one has to consider the correlation $\langle \sigma_i^z \sigma_{i+R}^z \rangle$ in the limit $R \to \infty$.
The difference between the results (\ref{ZZinf}) and (\ref{rhoz}) is attributed to an end-effect \cite{P1970}.
The result (\ref{rhoz}) leads directly to the  bulk magnetization $M_z$, 
with $\rho_z = M_z^2$ and thus $M_z = (1-\lambda^2)^{1/8}$ for $\lambda < 1$, with $M_z=0$ for $\lambda \ge1$.

\section{Exact results for $N =2$}
\label{results2}

Before calculating various spin correlations, we need to consider the solutions of equation (\ref{eqn}),  
which differ according to the physical regime described by $\lambda$.
For $\lambda=\lambda_c= 1$ the $L$ roots are given simply by
\begin{equation}
k_j = \frac{2j\pi}{2L+1}, \quad j=1,\ldots,L.
\label{soln}
\end{equation}
In contrast to the periodic case, more work is required for other values of $\lambda$.
There are $L$ real roots in the interval $(0,\pi)$ for $\lambda \ge 1+1/L$, while for 
$\lambda < 1+1/L$ there are $L-1$ roots in $(0,\pi)$, the remaining root being complex, of the form 
\begin{equation}
k_L = \pi + {\mathrm i} v,
\end{equation}
where $v$ satisfies the equation 
\begin{equation}
\sinh(L+1) v = \lambda^{-1} \sinh Lv.
\end{equation}
In particular, it follows that $\mathrm{e}^{v} \approx 1/\lambda$ for large $L$, which features in the calculations below.

On the other hand, for $\lambda >1$ the roots can be approximated for large $L$ by
\begin{equation}
k_j \approx \frac{\pi j}{L} - \frac{\lambda}{1+\lambda} \frac{ \pi j}{L^2}, \qquad j = 1, \ldots, L. 
\label{approx}
\end{equation}
This result follows by writing 
\begin{equation}
L k_j = \pi j - \pi \kappa_j + O\left( \frac1L \right) , \qquad j = 1, \ldots, L
\end{equation}
which gives
\begin{equation}
\cot(\pi \kappa_j) = \frac{\lambda^{-1} + \cos(\pi j/L)}{\sin(\pi j/L)}
\end{equation}
with solution
\begin{equation}
 \pi \kappa_j  = \frac{\pi j}{2L} + \tan^{-1} \left[ \frac{\lambda-1}{\lambda+1} 
 \tan \left( \frac{\pi j}{2L} \right) \right].
\end{equation}
This expansion mirrors that for the XY spin chain with open boundary conditions \cite{LSM1961}.
Here we have taken the $N=2$ value of the corresponding results for the free parafermion chain \cite{ABL2017}.

The ground-state energy per site in the  limit $L \to \infty$ is given simply in terms of the 
hypergeometric function $F(a,b;c;z) = \, _2 \!\, F_1(a,b;c;z)$ by 
\begin{equation}
e_0(\lambda)  = - F\left(-\frac12,-\frac12;1;\lambda^2\right). 
\label{eN2}
\end{equation}
This result is valid for $\lambda \le 1$, with $e_0(\lambda) = \lambda \, e_0(1/\lambda)$ for $\lambda > 1$.

\subsection{End-to-end correlations}

First consider the calculation of the correlations (\ref{XX}), (\ref{YY}) and (\ref{ZZ}) at $\lambda=1$,  
for which the relevant sums involving the eigenvector components with the exact roots (\ref{soln}) can be evaluated in closed form.
We have
\begin{eqnarray}
G_{1L}  &=& \frac{4}{2L+1} \sum_{j=1}^L (-1)^j \left[ \sin \left(\frac{2 \pi j L}{2L+1} \right) \right]^2   \cr
&=& \frac{(-1)^L + \cos (3L+1)\pi \, \sec \left( \frac{2L \pi}{2L+1} \right)}{2L+1} ,
\label{G1L}
\end{eqnarray}
\begin{eqnarray}
G_{L1}  &=& \frac{4}{2L+1} \sum_{j=1}^L (-1)^j \left[ \sin \left(\frac{2 \pi j}{2L+1} \right) \right]^2  \cr
&=& \frac{(-1)^L + \cos (L+3)\pi \, \sec \left( \frac{2 \pi}{2L+1} \right)}{2L+1} .
\label{GL1}
\end{eqnarray}
A similar result for the sum 
\begin{equation}
G_{11} =G_{LL} = \frac{4}{2L+1} \sum_{j=1}^L (-1)^j  \sin k_j \, \sin L k_j 
\end{equation}
is too cumbersome to reproduce here. 
Nevertheless, it can be readily established from the result that 
\begin{equation}
\lim_{L\to \infty} (G_{11} =G_{LL}) = - \frac{8}{3\pi}.
\end{equation}
As can be seen from the closed form expressions (\ref{G1L}) and (\ref{GL1}) 
both $G_{1L}$ and $G_{L1}$ vanish in this limit. 
Putting these results together gives
\begin{equation}
\rho_x = \left( \frac{8}{3\pi} \right)^2 \quad {\mathrm{at}} \quad \lambda=1 , 
\label{rhox1}
\end{equation}
with $\rho_z = 0 $ and $\rho_y = 0 $.

Turning to $\lambda < 1$,  the sum $G_{1L}$ is dominated with increasing $L$ by the complex root $k_L$, with 
\begin{equation}
G_{1L} \approx (A_{k_L}  \sin L k_L )^2 
\approx 1-\lambda^2 ,
\end{equation}
from which Pfeuty's result (\ref{ZZinf}) for $\rho_z$ follows.
On the other hand, $G_{L1}$ vanishes for large $L$, and so $\rho_y=0$ in this regime.
It follows that to calculate $\rho_x$ from (\ref{XX}), one needs to determine the value of $G_{11}=G_{LL}$. 
In the limit $L\to \infty$ 
\begin{equation}
G_{11} = G_{LL} = F\left(-\frac12,\frac12;2;\lambda^2\right), 
\end{equation}
giving the result
\begin{equation}
\rho_x(\lambda) = \lambda^2 \, F\left(-\frac12,\frac12;2;\lambda^2\right)^2 , \quad \lambda \le 1.
\label{XXinf}
\end{equation}
This result reduces to (\ref{rhox1}) at $\lambda=1$.

Unlike the end-to-end correlation $\rho_z$ given in (\ref{ZZinf}), 
the end-to-end correlation $\rho_x$ does not vanish for $\lambda > 1$. 
Rather the result for $\lambda > 1$ satisfies the duality relation
\begin{equation}
\rho_x({1}/{\lambda}) = \lambda^2  \rho_x(\lambda), 
\label{cXN2}
\end{equation}
with $0 \le \rho_x(\lambda) \le 1$.

\subsection{Nearest-neighbor correlations}

When periodic boundary conditions are applied, the end-to-end correlations are equivalent to nearest-neighbor correlations. 
For periodic boundary conditions, the results for various Ising correlations can be obtained   
from the key results given in McCoy's book \cite{McCoy2010}. 
Using different notation to distinguish the periodic case, the two nearest-neighbour correlations in the limit $L \to \infty$ are 
\begin{equation}
\langle \sigma_1^z \sigma_2^z \rangle   =  \begin{cases}
~~ F\left(-\frac12,\frac12;1;\lambda^2\right), \qquad ~ \lambda \le 1,\\
~\\
%\qquad \qquad \frac{2}{\pi}, \qquad  \qquad \quad \lambda=1,\\
%~\\
~ \frac{1}{2\lambda} F\left(\frac12,\frac12;2;\frac{1}{\lambda^2}\right), \quad \quad  \, \lambda \ge 1.
\end{cases}
\label{Zper}
\end{equation}
\begin{equation}
\langle \sigma_1^x \sigma_2^x \rangle   =  
\begin{cases}
\frac{\lambda^2}{4} F\left(\frac12,\frac12;2;\lambda^2\right)^2 
+\frac{\lambda^2}{8} F\left(-\frac12,\frac12;1;\lambda^2\right) F\left(\frac12,\frac32;3;\lambda^2\right), \quad  \lambda \le 1,\\
~\\
%\qquad \qquad \qquad \qquad \qquad \frac{16}{3\pi^2}, \qquad  \qquad \qquad  \qquad \qquad   \qquad ~~ \lambda=1,\\
%~\\
F\left(-\frac12,\frac12;1;\frac{1}{\lambda^2}\right)^2 
+\frac{1}{4\lambda^2} F\left(\frac12,\frac12;2;\frac{1}{\lambda^2}\right) F\left(-\frac12,\frac32;2;\frac{1}{\lambda^2}\right), ~~~  \lambda \ge 1.
\end{cases}
\label{Xper}
\end{equation}
At $\lambda=1$ these results reduce to the values 
\begin{equation}
\langle \sigma_1^z \sigma_2^z \rangle = \frac{2}{\pi}, \qquad \langle \sigma_1^x \sigma_2^x \rangle = \frac{16}{3\pi^2}. 
\label{per1}
\end{equation}

A related result  is the single site magnetization $M_x = \langle \sigma_1^x \rangle$, given by \cite{McCoy2010}
\begin{equation}
M_x =  
\begin{cases}
~~ \frac{\lambda}{2} F\left(\frac12,\frac12;2;\lambda^2\right), \qquad \qquad ~~ \, \lambda \le 1,\\
~\\
~~ F\left(-\frac12,\frac12;1;{\lambda^{-2}}\right), \qquad \qquad    \lambda \ge 1, 
\end{cases}
\label{Mx}
\end{equation}
with $M_x = \frac{2}{\pi}$ at $\lambda=1$. 
For systems with periodic boundary conditions the correlations 
$\langle \sigma_n^z \sigma_{n+1}^z \rangle$, $\langle \sigma_n^x \sigma_{n+1}^x \rangle$ 
and $M_x = \langle \sigma_n^x \rangle$ are independent of site $n$.
The Hellmann-Feynman theorem \cite{McCoy2010} can then be used to relate 
$\langle \sigma_n^z \sigma_{n+1}^z \rangle$ and $M_x$ to the ground-state energy $e_0(\lambda)$,  
with result
\begin{eqnarray}
\langle \sigma_n^x \rangle &=& - \frac{\partial e_0(\lambda)}{\partial \lambda},   \\
\langle \sigma_n^z \sigma_{n+1}^z \rangle &=&  -e_0(\lambda) + \lambda \frac{\partial e_0(\lambda)}{\partial \lambda}.
\end{eqnarray}
The various relations among these results can be checked by applying the relation 
\begin{equation}
\frac{\partial}{\partial z} F(a,b;c;z) =  \frac{a b}{c} F(a+1,b+1;c+1;z), \label{rel}
\end{equation}
among other identities \cite{McCoy2010,GR}.

On the other hand, no such translational invariance holds for systems with open boundary conditions.
For the Ising model, the nearest-neighbor correlations are site-dependent. 
At the boundary,  
\begin{eqnarray}
\langle \Psi_0 | \sigma^x_1 \sigma^x_2  | \Psi_0 \rangle &=& G_{11} G_{22} - G_{12} G_{21},  \label{XX12} \\
\langle \Psi_0 | \sigma^y_1 \sigma^y_2  | \Psi_0 \rangle &=& G_{21}, \label{YY12}\\
\langle \Psi_0 | \sigma^z_1 \sigma^z_2  | \Psi_0 \rangle &=& G_{12}, \label{ZZ12}
\end{eqnarray}
The limits of interest are
\begin{equation}
\rho^a_{12} = \lim_{L\to\infty} \langle \Psi_0 | \sigma^a_1 \sigma^a_2  | \Psi_0 \rangle,
\end{equation}
with $a=x,y,z$. 
These correlations do not appear to have been calculated explicitly for open boundary conditions.
They are straightforward to evaluate at $\lambda = 1$.
For this value the general nearest-neighbor correlations $\rho^z_{n,n+1} $ and $\rho^x_{n,n+1}$ are given by 
\begin{eqnarray}
\rho^z_{n,n+1} &=& \frac{8 \, n}{(4n+1)\, \pi}, \\
\rho^x_{n,n+1} &=& \frac{2048 \, n^2 (n+1)(2n+1)}{3(4n-1)(4n+3)(4n+1)^2 \, \pi^2}.
\end{eqnarray}
For $n=1$ these results simplify to $\rho^z_{12} =\frac{8}{5 \pi}$ and $\rho^x_{12} = \frac{4096}{525 \pi^2}$. 
The periodic values (\ref{per1}) are recovered in the limit $n \to \infty$. 
Physically, the periodic results are thus recovered for sites deep inside the open chain.

\section{Exact results for general $N$}
\label{exactN}

In this Section, in the absence of results for the eigenvectors of the $Z(N)$ free parafermion chain, 
we derive the ground-state expectation values which can be obtained by applying the 
Hellmann-Feynman theorem adapted for non-Hermitian systems (see, e.g., \cite{nonHQM}). 
The quantities considered are defined for finite-size chains by 
\begin{eqnarray}
&& \langle \Phi_0 | \, \tau_n   | \Psi_0 \rangle, \label{ttt} \\
&& \langle \Phi_0 | \,  \sigma_n \sigma_{n+1}^\dagger  | \Psi_0 \rangle . \label{sss}
\end{eqnarray}
For non-Hermitian systems, both the left $\langle \Phi |$ and right $| \Psi \rangle$ eigenvectors need to be considered.
For example, the ground-state energy $E_0$ is given by 
\begin{equation}
\langle \Phi_0 | E_0 = \langle \Phi_0 | H_{\mathrm{pf}} \qquad \mathrm{and} \qquad H_{\mathrm{pf}} | \Psi_0 \rangle = E_0 | \Psi_0 \rangle.
\end{equation}
The eigenvectors are not orthogonal, but rather form a biothonormal basis
\begin{equation}
\langle \Phi_m  | \Psi_n \rangle = \delta_{mn}, \qquad \sum_n | \Psi_n \rangle \langle \Phi_n | = 1.
\end{equation}
Moreover, for $N > 2$ the eigenvectors are no longer real. 
Nevertheless, the expectation values (\ref{ttt}) and (\ref{sss}) are observed to be real. 
As is the ground-state energy $E_0$.

More specifically, the quantities of interest here are defined by 
\begin{eqnarray}
\rho^X(\lambda) &=& \lim_{L\to\infty} \langle \Phi_0 | \, \tau_n  | \Psi_0 \rangle, \label{cXinf}\\
\rho^Z(\lambda) &=& \lim_{L\to\infty} \langle \Phi_0 | \, \sigma_n \sigma_{n+1}^\dagger  | \Psi_0 \rangle. \label{cZZinf}
\end{eqnarray}
In analogy with applying this approach for the Ising case with open boundary conditions, 
the values of $n$ are such that $1 \ll n \ll L$, i.e., for sites $n$ far from the ends of the chain.
We apply the Hellmann-Feynman theorem under this condition, 
and assuming that the contributions from sites near the boundaries become negligible compared to the contributions from the bulk.
The expected results are thus 
\begin{eqnarray}
\rho^X(\lambda) &=& - \frac{\partial e_0(\lambda)}{\partial \lambda},   \\
\rho^Z(\lambda) &=&  -e_0(\lambda) + \lambda \frac{\partial e_0(\lambda)}{\partial \lambda}.
\end{eqnarray}
We can now make use of the exact result for the ground-state energy per site, 
given in the $L \to \infty$ limit by \cite{ABL2017} 
\begin{equation}
e_0(\lambda)  = - F\left(-\frac1N,-\frac1N;1;\lambda^N\right).
\end{equation}
Like the $N=2$ result (\ref{eN2}), this result is valid for $\lambda \le 1$, with $e_0(\lambda) = \lambda \, e_0(1/\lambda)$ for $\lambda > 1$.

Subject to the above proviso for the values of $n$, and making use of the relation (\ref{rel}), the results for general $N$ follow in terms 
of hypergeometric functions as
\begin{equation}
\rho^X(\lambda) =  
\begin{cases}
~ \frac{1}{N} \lambda^{N-1} F\left(\frac{N-1}{N},\frac{N-1}{N};2;\lambda^N\right), \qquad  ~~~~ \lambda \le 1,\\
~\\
%F\left(-\frac1N,-\frac1N;1;\lambda^{-N} \right) 
%-\frac{1}{N} \lambda^{-N} F\left(\frac{N-1}{N},\frac{N-1}{N};2;\lambda^{-N} \right) , ~~~  \, \lambda \ge 1, 
~ F(-\frac1N, \frac{N-1}{N}; 1; \lambda^{-N}), \qquad \qquad \qquad  ~~ \lambda \ge 1,
\end{cases}
\label{Xgen}
\end{equation}
and
\begin{equation}
\rho^Z(\lambda)   =  \begin{cases}
~ F(-\frac1N, \frac{N-1}{N}; 1; \lambda^{N}), \qquad \qquad \qquad  ~~ \lambda \le 1,\\
%~ F\left(-\frac1N,-\frac1N;1;\lambda^{N} \right) 
%+\frac{1}{N} \lambda^{N} F\left(\frac{N-1}{N},\frac{N-1}{N};2;\lambda^{N} \right), \qquad ~ \lambda \le 1,\\
~\\
~ \frac{1}{N} \lambda^{1-N} F\left(\frac{N-1}{N},\frac{N-1}{N};2;\lambda^{-N}\right), \qquad     \lambda \ge 1.
\end{cases}
\label{Zgen}
\end{equation}
Here we have made use of the identity \cite{GR}
\begin{equation}
c \, F(a,b;c;z) - c \, F(a,b+1;c;z) + a z \, F(a+1,b+1;c+1;z) = 0.
\end{equation}

As to be expected, when $N=2$ the above results reduce to the Ising results (\ref{Mx}) and (\ref{Zper}) given in Section 3.2.
At $\lambda=1$ they simplify to the values
\begin{equation}
\rho^X = \rho^Z = \frac{1}{N} \, \frac{\Gamma(\frac{2}{N})}{\Gamma(\frac{N+1}{N})^2}  . 
\end{equation}
In general $\rho^X(\lambda) = \rho^Z(1/\lambda)$.

In the next Section we test the results (\ref{Xgen}) and (\ref{Zgen}) numerically.

\section{Numerical tests}
\label{resultsN}

In testing the above results numerically the left eigenvectors appearing in the expectation values (\ref{ttt}) and (\ref{sss}) 
can be calculated from $H_{\mathrm{pf}}^\dagger | \Phi_0 \rangle = E_0 | \Phi_0 \rangle$.
For this particular non-Hermitian model, this result is equivalent to that of a spatial symmetry observed by Baxter \cite{Baxter1989a}, see also \cite{Fendley2014}.
The numerical results are obtained using DMRG implemented by the ITensor library \cite{library}. 
Some modifications were required to support non-Hermitian systems; primarily, the Davidson eigensolver was replaced with Arnoldi. 
This results in a considerable slowdown, so the chain lengths typical for Hermitian systems could not be achieved. 
For this model we can take advantage of the ground-state energy being known exactly for finite size $L$, 
from which the difference with the numerical energy determined using DMRG was used as the convergence parameter. 
The numerical results presented here were converged to a tolerance of $10^{-6}$ relative to the exact result for $E_0$.

%%%%%%%%%%%%%%%%%%%%%%%%%%%%%%%%%%%%%%%
\begin{figure}[t]
\includegraphics[angle=0,width=1.0\textwidth] {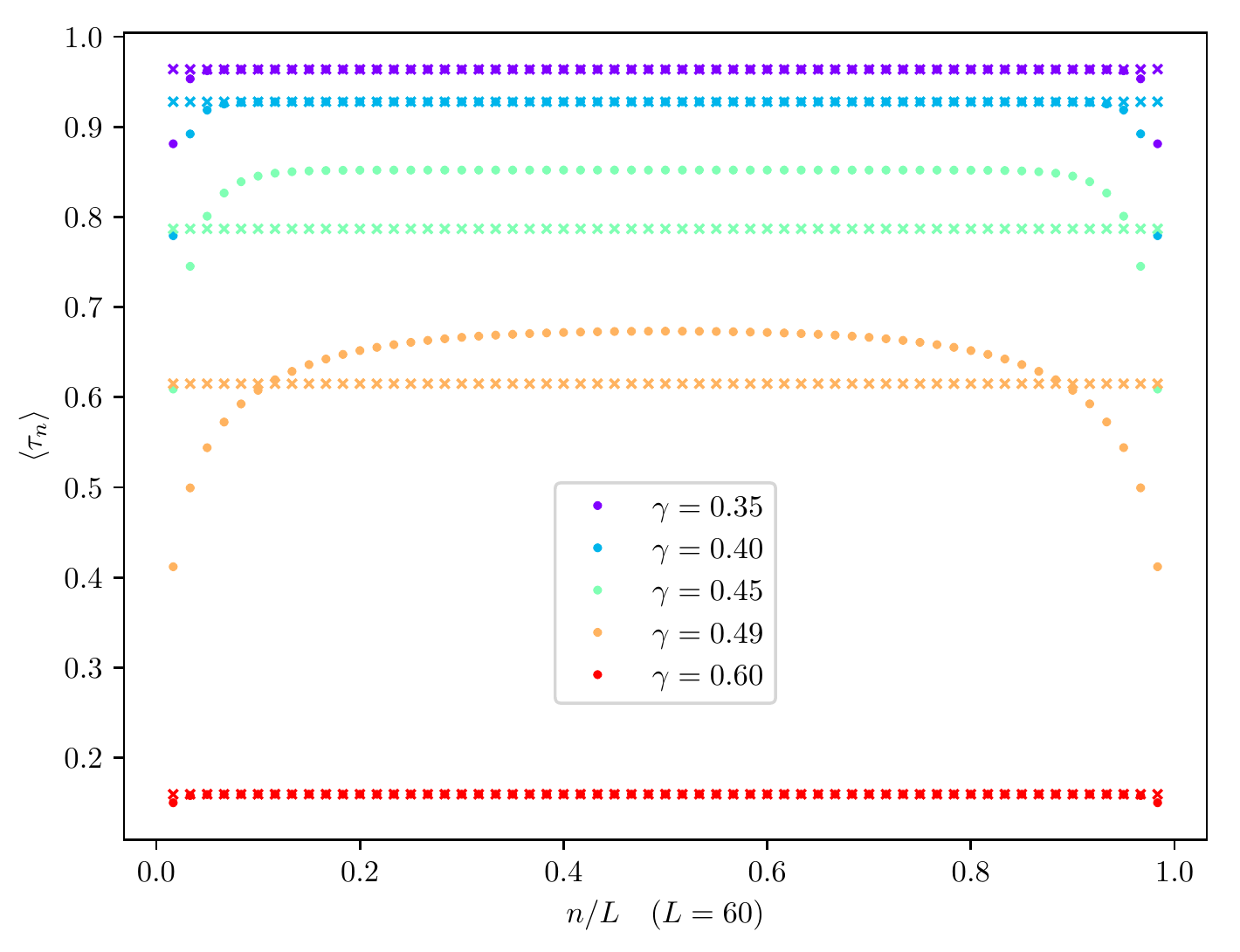}
\caption{
Site-dependence of the ground-state expectation value (\ref{ttt}) for the $Z(3)$ free parafermion spin chain for 
selected values of the coupling parameter $\gamma$. 
The values for open boundary conditions are indicated by circles. For comparison the corresponding values for periodic 
boundary conditions are indicated by crosses. 
}
\label{figure1}
\end{figure}
%%%%%%%%%%%%%%%%%%%%%%%%%%%%%%%%%%%%%%%

%%%%%%%%%%%%%%%%%%%%%%%%%%%%%%%%%%%%%%%
\begin{figure}[t]
\includegraphics[angle=0,width=1.0\textwidth] {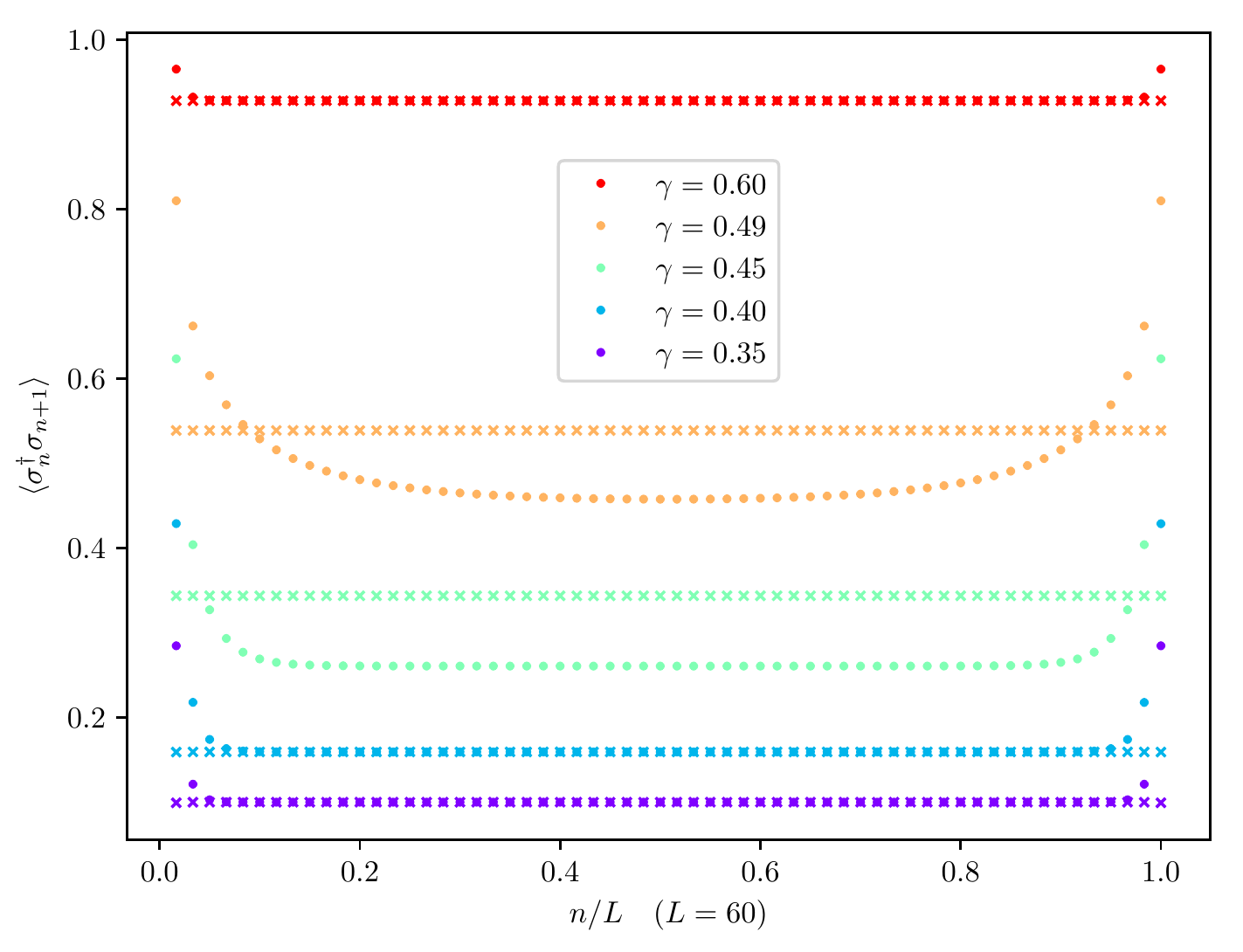}
\caption{
Site-dependence of the ground-state expectation value (\ref{sss}) for the $Z(3)$ free parafermion spin chain for 
selected values of the coupling parameter $\gamma$. 
The values for open boundary conditions are indicated by circles. For comparison the corresponding values for periodic 
boundary conditions are indicated by crosses. 

}
\label{figure2}
\end{figure}
%%%%%%%%%%%%%%%%%%%%%%%%%%%%%%%%%%%%%%%

Specifically, the ground-state energy of the $N^L \times N^L$ matrix $H_{\mathrm{pf}}$ is given by
\begin{equation}
E_0 = - \lambda \sum_{j=1}^L \epsilon_j^{1/N} .
\label{N0}
\end{equation}
In terms of the coupling parameter $g=1/\lambda^{N/2}$, the free parafermion quasi-energies $\epsilon_j$ 
appearing in (\ref{N0}) are the eigenvalues of the $L \times L$ matrix \cite{AB2018} 
\begin{equation}
\begin{bmatrix} 1 & g &   &  & &  \\ 
g &1+g^2 & g &&&\\   
%& &  &  &  & \\ 
&  & \ddots & \ddots &  \ddots & \\
%& &  &  &  & \\
& &  & g & 1+g^2 &g  \\
& &  &  & g & 1+g^2
\end{bmatrix}. 
\end{equation}

As already mentioned, we directly diagonalise this smaller matrix numerically to test the convergence. 
Alternatively, the value $E_0$ used in the convergence test 
may be obtained by numerically solving an equation for the quasi-momenta values determining $\epsilon_j$ \cite{ABL2017}.
In the numerical computations it is convenient to work in terms of the coupling parameter
\begin{equation}
\gamma =  \frac{1}{1+\lambda}.
\label{gamma}
\end{equation}
The critical point $\lambda_c =1$ thus corresponds to $\gamma_c=\frac12$, with the range of coupling parameters now restricted to $0 \le \gamma \le 1$.

Using DMRG as described above, the eigenvectors $\langle \Phi_0 |$ and $| \Psi_0 \rangle$ of $H_{\mathrm{pf}}$ are constructed for finite $L$ to 
evaluate the ground-state expectation values (\ref{ttt}) and (\ref{sss}).
These quantities are shown in figures 1 and 2 for $N=3$ and $L=60$ with site $n$ varying over the length of the chain. 
Very similar behaviour is observed for the $N=2$ Ising case.
In particular, away from the ends of the chain, the curves flatten, indicative of a `bulk' limiting result.
For the Ising case, the flattened curves agree with the values obtained for periodic boundary conditions.
For comparison, figures 1 and 2 also show the corresponding values obtained for periodic boundary conditions.
It is clear that the expectation values depend on the boundary conditions for $N=3$, 
consistent with the behaviour observed for the ground state energy \cite{AB2018}.
However, sufficiently away from criticality it appears that these values are no longer dependent on the boundary conditions. 
Convergence to the exact results (\ref{Xgen}) and (\ref{Zgen}) for $N=3$ can be seen in figures 3 and 4, 
which plot the ground-state expectation values at the centre of the open chain for increasing chain size $L$.
Similar convergence is observed for the $N=2$ Ising case for same size chains. 

%%%%%%%%%%%%%%%%%%%%%%%%%%%%%%%%%%%%%%%
\begin{figure}[t]
\includegraphics[angle=0,width=1.0\textwidth] {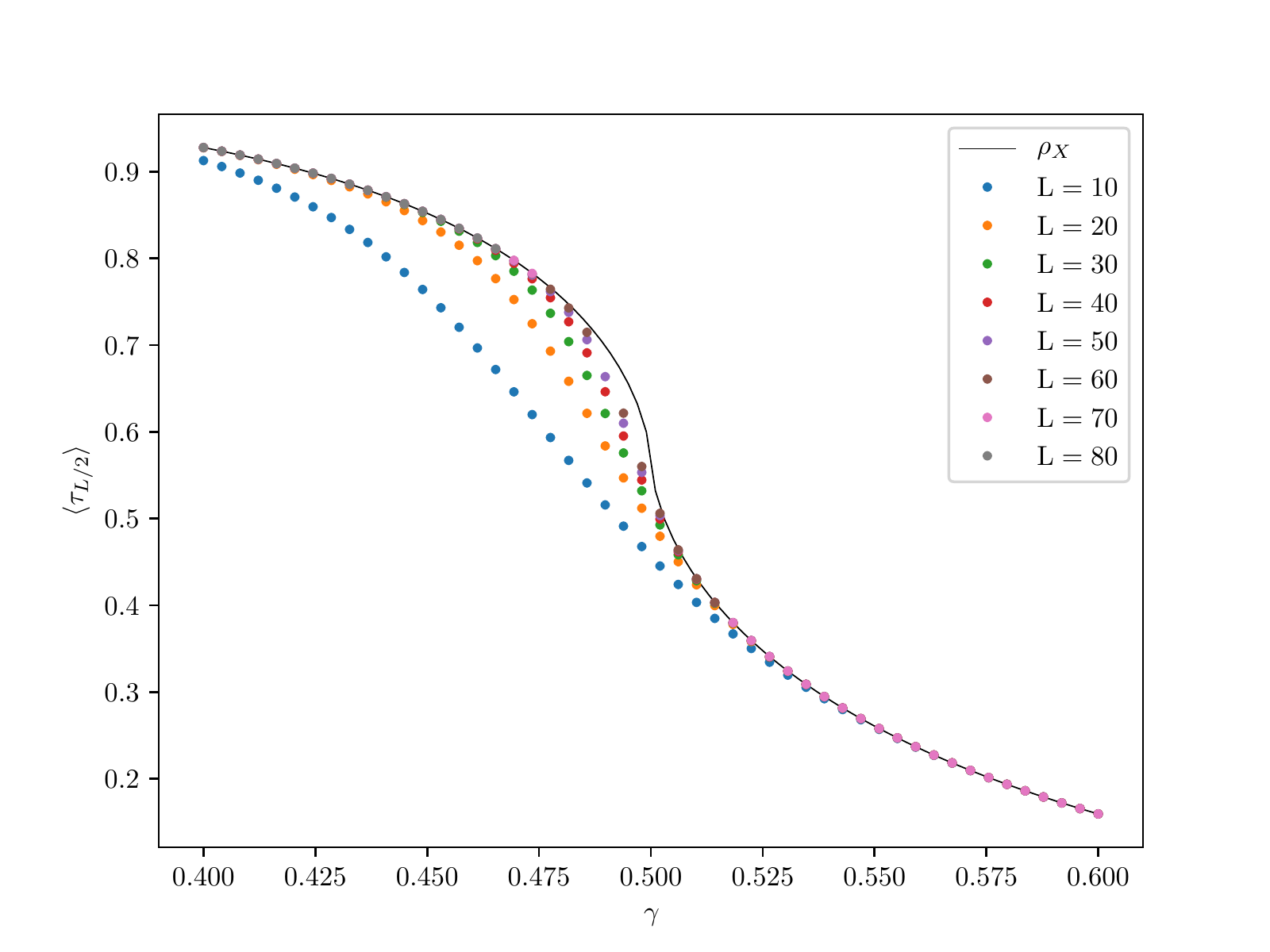}
\caption{
Ground-state expectation value (\ref{ttt}) for the midpoint of the $Z(3)$ free parafermion spin chain with 
increasing chain size $L$. The solid line is the exact result (\ref{Xgen}).
}
\label{figure3}
\end{figure}
%%%%%%%%%%%%%%%%%%%%%%%%%%%%%%%%%%%%%%%

%%%%%%%%%%%%%%%%%%%%%%%%%%%%%%%%%%%%%%%
\begin{figure}[t]
\includegraphics[angle=0,width=1.0\textwidth] {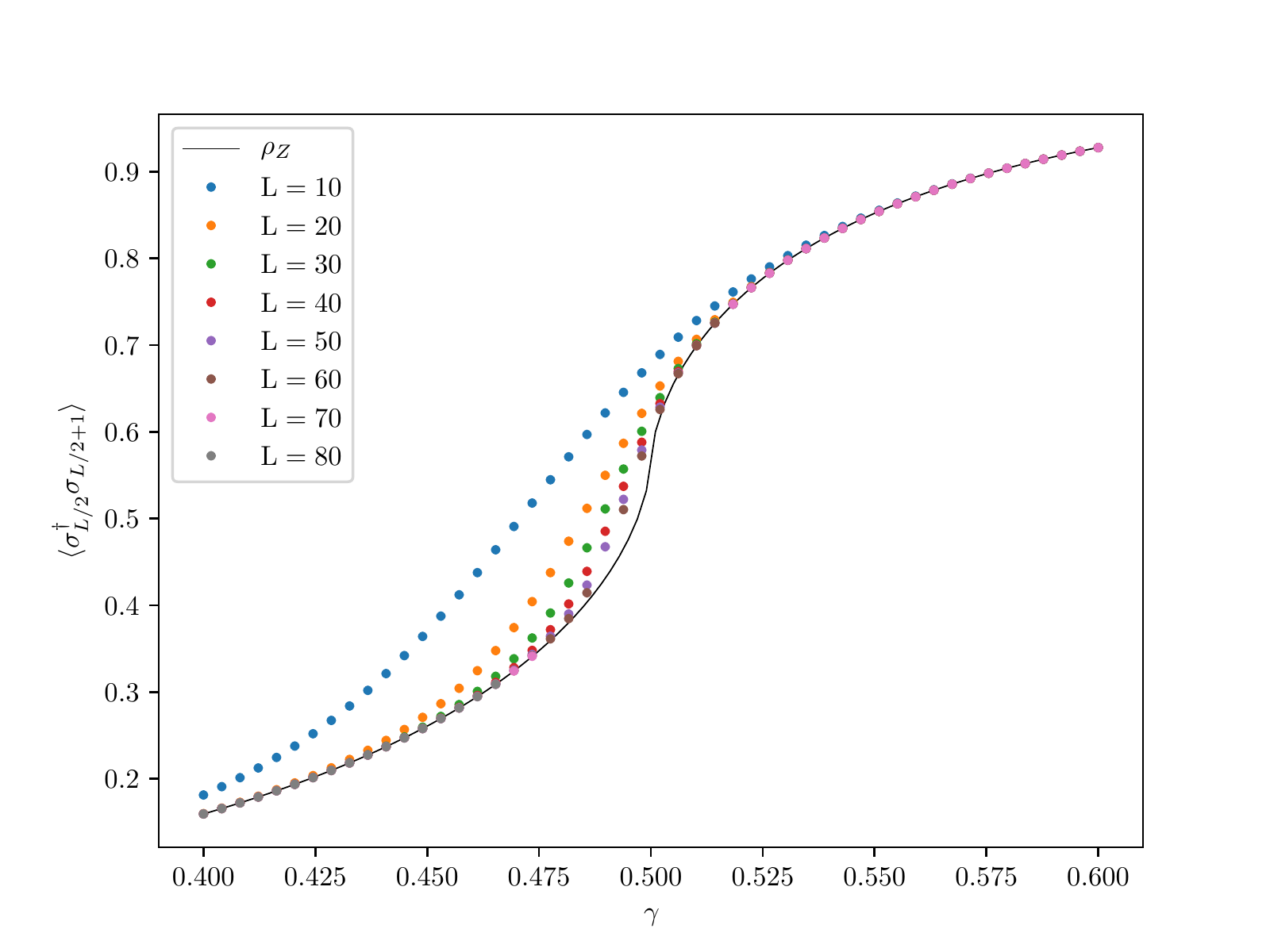}
\caption{
Ground-state expectation value (\ref{sss}) for the midpoint of the $Z(3)$ free parafermion spin chain with 
increasing chain size $L$. The solid line is the exact result (\ref{Zgen}).
}
\label{figure4}
\end{figure}
%%%%%%%%%%%%%%%%%%%%%%%%%%%%%%%%%%%%%%%

\section{Discussion}
\label{conclude}

Motivated by the more general free parafermion $Z(N)$ model (\ref{pf}),  
we have studied elementary correlations for the quantum Ising model (\ref{ham}) subject to open boundary conditions. 
The  results for the Ising model have been given in Section 3.
Our aim has been to generalise these results to the $Z(N)$ model.
We have been able to do this in Section 4 for the particular ground-state expectation values obtained by applying the Hellmann-Feynman theorem.
The key results are given in equations (\ref{Xgen}) and (\ref{Zgen}). 
We tested these results numerically in Section 5 for $N=3$ and believe them to be correct for general $N$.

As tempting as it may be,  we have not referred to these results as correlations. 
This is because in general non-Hermitian systems are non-unitary, leading to the problem of formulating a meaningful quantum mechanical description with 
physical correlations. 
The well known exception is the class of non-Hermitian systems which are ${\cal P}{\cal T}$ symmetric, with a real eigenspectrum for which  
a definite metric is guaranteed \cite{Binder}. 
Here one can also define quasi-Hermitian and pseudo-Hermitian systems. 
For both types the eigenvalues are real, but no definite conclusions can be made with regard to the existence of a definite metric for the latter.

Of further interest are the Ising model end-to-end correlations (\ref{ZZinf}) and (\ref{XXinf}).
The simple result (\ref{ZZinf}) was obtained long ago by Pfeuty \cite{P1970}.
The corresponding end-to-end correlation $\rho_x$ given by (\ref{XXinf}) does not appear to have been discussed before. 
Rather than generalising in a relatively simple way for the $Z(N)$ model, like for the order parameters (\ref{spon}) 
or the results (\ref{Xgen}) and (\ref{Zgen}) presented here, such end-to-end `correlations' are seen to
diverge with increasing system size at the critical value $\lambda_c=1$. 
In contrast to Hermitian systems, such divergent behaviour should not be surprising for the physics of non-Hermitian systems. 
An example from optics is the behaviour of the Petermann factor, 
which depends on the overlap between the left and right eigenvectors of a non-unitary wave operator. 
This factor takes very large values at resonances which can be explained by degeneracies in the eigenspectrum \cite{Berry1}. 
Another example relates to optical singularities in non-Hermitian chiral crystals \cite{Berry2}. 
This divergent behaviour will be discussed in detail elsewhere.

We remain optimistic that it should be possible to derive exact results for various `correlations' of the $Z(N)$ model described by free parafermions, 
including the results obtained here via the Hellmann-Feynman theorem.
Recalling the summary of the free fermion formalism given in Section 2, 
it is possible to obtain analogous eigenvector components $\psi$ and $\phi$ for the $Z(N)$ model via the corresponding $L \times L$ matrices $C$ and $C^\dagger$ \cite{AB2018}.
However, there is a conceptual difficulty in relating them back to the spin `correlations' of interest for the $Z(N)$ model. 
This is related to the problem of establishing (or knowing) an analog of Wick expansion for parafermions.
The Wick expansion is the key ingredient in the calculation of correlations in the free fermion XY and Ising chains \cite{LSM1961,K2004}.
Whether or not such a crucial step can be made for free parafermions, allowing the exact calculation of spin `correlations' and dynamics, 
remains to be seen.
There is a glimmer of hope, however.
This relates to the observation that the $Z(N)$ model in question satisfies the property of reflection positivity \cite{Jaffe2015}.
Reflection positivity has been noted as a perturbative form of Wick rotation.

In concluding,  we remark that after all these years it remains a major challenge to calculate the eigenvectors of the $N$-state chiral Potts model. 
Some development was made in calculating the eigenvectors for the superintegrable case \cite{AYP2008, AYP2009, ND2008, R2010},  
but not sufficiently to allow the explicit calculation of correlation functions, and for example the calculation of the order parameters (\ref{spon}) via this approach. 
In comparison it would seem that the exact calculation of the eigenvectors of the free parafermion $Z(N)$ model may provide an easier challenge. 
Indeed, we note that some results have been reported for the eigenvectors of this model \cite{IST2008}, but the results are far from transparent. 
We hope that the progress reported here will inspire further work on this model.

\ack
We are grateful to Francisco Alcaraz, Rodney Baxter, Michael Berry, Fabian Essler and Paul Fendley for helpful comments at various stages of this work.
This project has been supported by the Australian Research Council Discovery Project DP180101040 and 
the National Natural Science Foundation of China Grant No.~11575037.

%%%%%%%%%%%%%%%%%%%%%%%%%%%%%%%%%%%%%%%%%%%%%%%%%%%%%%%%%%%%%%%%%%%%%%%%%%%%%%%%%
\section*{References}
%%%%%%%%%%%%%%%%%%%%%%%%%%%%%%%%%%%%%%%%%%%%%%%%%%%%%%%%%%%%%%%%%%%%%%%%%%%%%%%%%

\providecommand{\newblock}{}

\end{document}